\documentclass[lettersize,journal]{IEEEtran}
\usepackage{amsmath,amsfonts}
\usepackage{amssymb}
\usepackage{mathtools}
\usepackage{algorithmic}
\usepackage{mathrsfs}
\usepackage{booktabs}
\usepackage{makecell}
\usepackage{algorithm}
\usepackage{array}
\usepackage[caption=false,font=normalsize,labelfont=sf,textfont=sf]{subfig}
\usepackage{textcomp}
\usepackage{color}
\usepackage{stfloats}
\usepackage{url}
\usepackage{verbatim}
\usepackage{xcolor}
\usepackage{graphicx}
\usepackage{cite}

\usepackage[T1]{fontenc}
\usepackage{mathptmx}
\usepackage{amsmath,amssymb,amsthm,bm}
\usepackage{microtype}
\usepackage[hidelinks]{hyperref}
\newtheorem{proposition}{Proposition}

\begin{document}

\title{
Loadability Assessment for Large-Load Integration:
From Repeated Power Flow Tracing to Historical Trajectory Transfer
}

\author{Yihan Huang,~\IEEEmembership{Student Member,~IEEE}, Yang Weng,~\IEEEmembership{Senior Member,~IEEE}


\thanks{Yihan Huang and Yang Weng are with the School of Electrical, Computer and Energy Engineering, Arizona State University, Tempe, AZ 85281 USA (e-mail: yhuan643@asu.edu; yang.weng@asu.edu)}
}

\markboth{}%
{Shell \MakeLowercase{\textit{et al.}}: A Sample Article Using IEEEtran.cls for IEEE Journals}

\IEEEpubid{}

\maketitle

\begin{abstract}
Integrating large data-center loads requires utilities
to evaluate voltage-stability loading margins under
different generation and demand conditions.
Continuation power flow (CPF) traces P--V curves
through successive prediction and correction steps,
but related assessments can repeatedly reconstruct
similar nonlinear behavior. Reusing an operating
point as an initial guess does not capture the
subsequent voltage response. This paper proposes
a continuation method that transfers finite segments
of historical power flow (PF) trajectories.
Each segment is aligned with the present state
and continuation direction, supplementing the CPF
tangent predictor with previously computed nonlinear
curve shape. Short Newton tests evaluate compatibility
with the present PF equations, and an attention-based
selector combines their results with network conditions,
loading patterns, and recent continuation states.
Numerical correction and fold refinement determine
accepted solutions and the loading limit.
The analysis bounds prediction error through
differences in trajectory curvature and establishes
sufficient conditions for local correction and
branch preservation. Studies on modified IEEE
14- and 118-bus systems under renewable-generation
and data-center load variations demonstrate reduced
online computation and higher observed recovery
of specified continuation solutions than Adaptive CPF.
For loading-limit assessment, paired online
speedups are 4.88--21.62$\times$ over Adaptive CPF
and 1.53--4.77$\times$ over holomorphic embedding
with arc-length parametrization (HEAP) across 
the tested scenarios.
\end{abstract}


\section{Introduction}

\IEEEPARstart{I}{ntegrating} large data-center loads
requires utilities to evaluate voltage-stability loading
margins under different generation and demand conditions.
Data centers are becoming a major driver of electricity
demand growth \cite{Xiao2025AIServers,Gyang2026DataCenter},
while increasing renewable generation
introduces further variation in network injections
\cite{Luderer2022Electrification}.
The unexpected disconnection of nearly 4,000 MW of
data-center load in northern Virginia in July 2026
illustrates the scale of concentrated computational
demand \cite{PJM2026LargeLoad}.
Alongside disturbance-response studies, utilities must
determine how much additional demand a network can
support before reaching a voltage-stability loading
limit. Changes in proposed loads and renewable generation
require this assessment to be repeated across operating scenarios
\cite{Sun2023StaticVoltage,Wu2022TriSectional}.

The problem considered here is how to use historical 
PF trajectories to reduce repeated tracing
of related P--V curves. Different injections and
loading patterns can produce curves with similar
nonlinear shapes but different voltage profiles
and loading limits. A historical operating point
provides an initial guess without describing the
subsequent voltage response, and similar starting
voltages do not ensure similar evolution.
Near a voltage-stability nose, multiple PF solutions
can also correspond to the same loading level
\cite{Ajjarapu1992CPF}. Useful historical information
must therefore support the specified advance along
the tracked branch, while the present PF equations
determine the corrected solution and loading limit.

Established numerical methods provide the foundation
for these assessments. Newton-based and fast-decoupled
methods compute operating points for specified injections
\cite{Tinney1967Newton,Stott1974FDLF}, while continuation
power flow (CPF) and CPFLOW trace connected solution
branches \cite{Ajjarapu1992CPF,Chiang1995CPFLOW}.
Higher-order predictors and holomorphic methods improve
trajectory approximation and limiting-point computation
\cite{lundberg1991,Zhang2022HEAP,Zhang2025RobustHE}.
Related methods characterize admissible
renewable-injection regions
\cite{Zeng2024ADMIRE,Zeng2026CoreX}.
Multi-dimensional holomorphic embedding supports
scenario reuse through a prepared representation
in selected injection variables
\cite{liu2017,Sun2023StaticVoltage} and loading-margin
sensitivities estimate changes in a known limit
\cite{greene1997}, which provide different
forms of approximation and reuse. A prepared embedding
depends on its represented variables, and local margin
updates do not directly supply a finite trajectory
under a larger change. The opportunity pursued here
is to transfer previously computed P--V shape to
reduce new continuation work.

Learning-based approaches reduce repeated computation
through mappings between operating inputs and solutions.
Physics-guided neural networks and support-matrix
regression incorporate physical and network structure
into PF prediction
\cite{Hu2021PhysicsGuidedPF,Yuan2022SupportMatrixPF}.
Related optimal power flow (OPF) approaches consider
topology changes \cite{Zhou2023DeepOPFFT},
initialization information for distinguishing
solver-dependent solutions \cite{DeepOPFAL2023},
and unsupervised training with objectives and
physical constraints \cite{Huang2024UnsupervisedOPF}.
These advances demonstrate the value of physical
information and solution context. Nevertheless,
predicting an operating point or initial guess
differs from supplying the nonlinear evolution
leading to a specified continuation solution.
Changes beyond the training distribution further
challenge learned mappings
\cite{DomainGeneralizationSurvey}. For loadability
assessment, the relevant information includes how
voltages and loading evolve along the tracked branch.

The main idea of this paper is to transfer that
evolution through a finite historical PF segment.
Matching its starting state and tangent to the
present continuation state separates differences
in operating level and initial direction from
the nonlinear shape available for reuse.
The transferred endpoint equals the present CPF
tangent prediction plus the aligned historical
segment's departure from its own tangent prediction.
Previously computed curve bending therefore
supplements the local predictor, and the historical
ending tangent supports intermediate predictions.

Historical PF trajectory can nevertheless be misleading
under changed injections. The proposed method
therefore evaluates candidate segments through short
Newton correction tests using the present PF equations.
Power mismatches, mismatch reduction, and directional 
agreement provide evidence for an attention-based selector, 
together with network conditions, loading patterns, 
and recently accepted continuation states. The selected 
segment guides numerical prediction and correction, 
with step reduction or ordinary CPF when guidance is unsuitable. 
Final refinement solves the present PF and fold
equations. The analysis bounds prediction error
through differences in historical and present
curve bending and establishes sufficient conditions
for local correction and branch preservation.

Numerical studies use a two-bus illustrative system, 
modified IEEE 14- and 118-bus networks under
renewable-generation and data-center load variations.
Continuation-target tests evaluate recovery of
specified branch solutions near and across noses.
Separate loading-limit studies compare the first-nose
assessment time with Adaptive CPF, holomorphic
embedding with arc-length parametrization (HEAP)
\cite{Zhang2022HEAP}, and the enhanced holomorphic embedding with arc-length parametrization (E-HEAP)
\cite{Zhang2025RobustHE}.
Across the evaluated loading-limit scenarios,
the proposed method achieves paired online speedups
of $4.88$--$21.62\times$ over Adaptive CPF and
$1.53$--$4.77\times$ over HEAP while passing the
common acceptance checks. Component comparisons
show reduced work from historical reuse and alignment,
but do not establish an independent population-level
recovery gain from Newton testing alone.
Reported online costs include candidate testing,
correction, and fallback; offline training and
database construction are excluded.

The rest of the paper is organized as follows.
Section~\ref{sec:problem} formulates the problem.
Section~\ref{sec:method} presents historical trajectory
transfer and guided continuation.
Section~\ref{sec:theory} provides the theoretical
guarantees.
Section~\ref{sec:experimental_results} presents numerical
comparisons and component evaluations.
Section~\ref{sec:conclusion} concludes the paper.

\section{Problem Formulation}
\label{sec:problem}

Consider repeated voltage-stability loadability assessments
for large-load integration under changing demand and
renewable generation. Each assessment specifies base
net injections and a loading pattern describing additional
demand and associated generation changes. Starting from
a feasible high-voltage operating point, the objective
is to determine the first loading maximum on the
connected PF solution branch.

Let
$\bm x=[\bm\theta^{\mathsf T},\bm V^{\mathsf T}]^{\mathsf T}$
contain non-slack voltage angles and PQ-bus voltage
magnitudes. The reduced PF equations are
\begin{equation}
 \bm F(\bm x,\lambda)=
 \begin{bmatrix}
 \bm P(\bm x)-\bm P^0-\lambda\Delta\bm P\\
 \bm Q(\bm x)-\bm Q^0-\lambda\Delta\bm Q
 \end{bmatrix}
 =\bm0,
 \label{eq:pf}
\end{equation}
where active-power equations apply to non-slack buses
and reactive-power equations apply to PQ buses.
The vectors $\bm P^0$ and $\bm Q^0$ are base net
injections, and $\Delta\bm P$ and $\Delta\bm Q$
specify their changes per unit loading parameter
$\lambda$. The base operating point has $\lambda=0$.
An increase in net demand corresponds to a negative
injection increment. The location and power factor
of additional demand, together with generation
participation, determine the loading pattern.
Prescribed voltage magnitudes at PV and slack buses
are PF model parameters.

Continuation treats $\lambda$ as an additional unknown,
allowing the solution curve to be traced through
simple folds where the ordinary PF Jacobian becomes
singular \cite{Ajjarapu1992CPF}. To account for the
different numerical scales of angles, voltage
magnitudes, and loading, define
\begin{equation}
 \bm z=\bm D^{-1}
 \begin{bmatrix}\bm x\\ \lambda\end{bmatrix},
 \label{eq:scaled-state}
\end{equation}
where $\bm D$ is a fixed positive diagonal matrix
shared by historical and present cases.
Angles are in radians, voltage magnitudes and powers
are in per unit, and $\lambda$ is dimensionless.
Below, $\bm F(\bm z)$ denotes \eqref{eq:pf}
evaluated at the variables recovered from
\eqref{eq:scaled-state}.

Let $s$ denote arc length in the scaled coordinates. The components of $\bm D\bm t$ give $d\bm\theta/ds$, $d\bm V/ds$, and $d\lambda/ds$. The tangent initially points toward increasing
loading and is subsequently oriented consistently
with the preceding accepted tangent.
The unit tangent $\bm t=d\bm z/ds$ satisfies
\begin{equation}
 \begin{aligned}
 \bm J_a\bm t=\bm0,\quad \|\bm t\|_2=1, \quad
 \bm J_a=
 \left[
 \bm J\ \ \frac{\partial\bm F}{\partial\lambda}
 \right]\bm D,\quad
 \bm J=\frac{\partial\bm F}{\partial\bm x}.
 \end{aligned}
 \label{eq:tangent}
\end{equation}

The loading limit and remaining margin for the PF nose are
\begin{equation}
 \begin{aligned}
 \lambda_{\rm lim}=\lambda(s_*),\
 \lambda'(s_*)=0,\ \lambda''(s_*)<0, \
 \Delta\lambda&=\lambda_{\rm lim}-\lambda_0,
 \end{aligned}
 \label{eq:nose}
\end{equation}
where $s_*$ is the first loading maximum reached
from the base operating point and $\lambda_0$
is the loading level at which the remaining margin
is evaluated. The MW margin is $\Delta\lambda$
times the aggregate demand increase per unit
$\lambda$, converted using the system power base.

The formulation considers smooth PF segments with
fixed topology, bus types, and control equations.
Continuation across reactive-power limits and
other control events is left for future work.
Historical segments used for direct alignment
share the present case's state dimension, bus
ordering, and compatible control configuration.
Different-sized networks require an additional
state mapping and are outside this formulation.

At each continuation step, the available information consists of a corrected state $\bm z_0$, its forward tangent $\bm t_0$, a short history of accepted continuation states and directions, a requested arc-length advance $h$, and a database of previously corrected PF segments. The requested advance is along the scaled PF curve; it does not prescribe a fixed increase in $\lambda$. Historical segments may end before their own loading maxima. Future states and the loading limit of the present case are unavailable online. The computational objective is to reuse these segments to reduce continuation work while retaining the specified PF accuracy, branch-tracking, and completion criteria.

\section{Historical Trajectory Transfer for Power Flow Continuation}
\label{sec:method}

Related loadability assessments can repeatedly reconstruct
similar nonlinear P--V behavior under changing injections
and loading patterns. The proposed method transfers this
behavior through finite historical PF segments. Each
candidate is aligned with the current corrected state
and continuation direction, tested using the present
PF equations, and evaluated as guidance for numerical
prediction and correction. Historical trajectories
supply previously computed curve shape, while the present
equations determine accepted states and the loading limit.

\subsection{Historical Segments and Prediction Information}

At the current corrected state $\bm z_0$, conventional CPF predicts the next solution using the local tangent as $\bm z_0+h\bm t_0$, where $h$ is the requested arc-length advance in the scaled coordinates of \eqref{eq:scaled-state}. This predictor captures the instantaneous voltage and loading response but does not account for changes in trajectory direction over a finite advance. Such changes become particularly important near a P--V nose, where $d\lambda/ds$ approaches zero while the voltage states continue to evolve. Consequently, local tangent prediction may require frequent corrections to follow strongly curved solution branches.

Historical PF trajectories provide additional information about this finite-distance evolution. Unlike an isolated operating point, a historical segment contains both the accumulated displacement and the change in tangent direction along the solution branch. These geometric quantities describe how the trajectory departs from its initial tangent prediction and how its direction evolves. When historical and present trajectories exhibit similar aligned shapes, this information can complement the local CPF predictor, even if their original voltage profiles and loading levels differ.

To exploit this information, we construct a historical database containing corrected PF states, unit tangents, segment lengths, and operating conditions. The stored segments share the state dimensions, bus ordering, and scaling defined in Section~\ref{sec:problem}, but need not extend to their own loading limits. At each continuation step, a small group of candidates is retrieved using the available operating conditions and trajectory descriptors. These candidates are subsequently aligned with the current state and continuation direction, before their compatibility is evaluated using the present PF equations.

\subsection{Transferring Nonlinear P--V Shape}

Let candidate $m$ be a historical segment
$\bm z_m(s)$, $0\leq s\leq h_m$, parameterized
by scaled arc length, with unit tangent $\bm t_m(s)$.
Its initial voltage profile, loading level, and
direction generally differ from the present case.
Define
\begin{equation}
 \alpha_m=\frac{h}{h_m},\qquad
 \bm R_m^{\mathsf T}\bm R_m=\bm I,\qquad
 \bm R_m\bm t_m(0)=\bm t_0.
 \label{eq:align}
\end{equation}
The aligned prediction over $0\leq s\leq h$ is
\begin{equation}
 \widetilde{\bm z}_m(s)=
 \bm z_0+\alpha_m\bm R_m
 \left[
 \bm z_m(s/\alpha_m)-\bm z_m(0)
 \right],
 \label{eq:transfer}
\end{equation}
with endpoint and ending tangent
\begin{equation}
 \widehat{\bm z}_m=\widetilde{\bm z}_m(h),\qquad
 \widehat{\bm t}_m=\bm R_m\bm t_m(h_m).
 \label{eq:endpoint}
\end{equation}
Translation matches the current state, rotation matches
the starting direction, and $\alpha_m$ adjusts the
segment length. Consequently,
$\widetilde{\bm z}_m(0)=\bm z_0$ and
$\widetilde{\bm z}'_m(0)=\bm t_0$.

The endpoint can equivalently be written as
\begin{equation}
 \widehat{\bm z}_m
 ={}\bm z_0+h\bm t_0 \notag +\alpha_m\bm R_m
 \left[
 \bm z_m(h_m)-\bm z_m(0)-h_m\bm t_m(0)
 \right].
 \label{eq:bending-displacement}    
\end{equation}
The bracketed term is the historical endpoint's
departure from its tangent prediction.
Thus the transferred prediction supplements the
present CPF tangent advance with historical nonlinear
shape. This term vanishes for a straight segment.
For a compatible curved segment, it supplies
finite-distance information about the evolving
voltage--loading response. The ending tangent
provides the corresponding predicted direction.
A rotation in the plane of the two starting tangents
is obtained by setting $\bm u=\bm t_m(0)$,
$\bm v=\bm t_0$, $c=\bm u^{\mathsf T}\bm v$, and
\begin{equation}
 \bm K=\bm v\bm u^{\mathsf T}-\bm u\bm v^{\mathsf T},
 \qquad
 \bm R_m=\bm I+\bm K+\frac{\bm K^2}{1+c}.
 \label{eq:rotation}
\end{equation}
Candidates with nearly opposite starting directions are excluded to avoid numerical instability. The rotation is applied only to construct predictions in scaled voltage--loading coordinates, without modifying the present-case PF equations. Since geometric alignment alone does not guarantee power balance or branch consistency, each prediction is tested using short Newton corrections under the present operating conditions.

\subsection{Testing Electrical Compatibility}

Each candidate receives up to limited Newton test
iterations, with early termination when screening
criteria fail. Starting from
$\bm z_m^{(0)}=\widehat{\bm z}_m$, solve
\begin{equation}
 \begin{bmatrix}
 \bm J_a(\bm z_m^{(j)})\\
 \widehat{\bm t}_m^{\mathsf T}
 \end{bmatrix}
 \Delta\bm z_m^{(j)}
 =
 -\begin{bmatrix}
 \bm F(\bm z_m^{(j)})\\
 \widehat{\bm t}_m^{\mathsf T}
       (\bm z_m^{(j)}-\widehat{\bm z}_m)
 \end{bmatrix},
 \label{eq:probe}
\end{equation}
and update
$\bm z_m^{(j+1)}
=\bm z_m^{(j)}+\Delta\bm z_m^{(j)}$.
All residuals and Jacobians use the present PF model.
The last row defines a continuation plane through
the predicted endpoint, perpendicular to its proposed
tangent. Both voltage variables and $\lambda$ can
change during correction. A suitable plane can keep
the bordered system nonsingular near a simple nose,
where the ordinary PF Jacobian becomes singular. 

The tests provide PF mismatches, correction magnitudes,
mismatch reduction, and directional agreement,
collected in the candidate feature vector $\bm e_m$.
Failed corrections, inadequate mismatch reduction,
or inconsistent directions make a candidate ineligible.
These tests identify unsuitable predictions without
fully tracing every candidate. The selected guide
remains subject to intermediate correction and
branch-tracking checks.

\subsection{Selecting Guidance Using Network \& Loading Information}

Candidates with similar endpoint mismatches can require
different subsequent continuation work. Selection
therefore combines their numerical test results with
the present loading pattern and recent voltage response.
Let $\mathcal E$ denote the eligible candidate set. The trajectory context is encoded using a topology-aware Transformer with temporal attention over accepted continuation states and bus-admittance-matrix-biased spatial attention across buses. The available information is encoded as
\begin{equation}
 \bm\chi=f_{\phi_c}
 \left(
 \bm\xi^{\rm bus},\bm\xi^{\rm br},\mathcal Z_{\rm acc}
 \right),
 \label{eq:encoder}
\end{equation}
where $\phi_c$ contains the encoder parameters.
Bus features $\bm\xi^{\rm bus}$ describe injections,
loading increments, voltages, and bus types.
Branch features $\bm\xi^{\rm br}$ describe connectivity
and electrical parameters. The ordered history
$\mathcal Z_{\rm acc}$ contains recently accepted
states and their continuation directions.
Only information available at the current step
enters the encoder.

The candidate scores and selection weights are
\begin{align}
a_m&=a_m^{\rm phys}+g_{\phi_s}\bigl(\bm\chi,\bm e_m,\{\bm e_j\}_{j\in\mathcal E}\bigr),\notag\\
w_m&=\frac{e^{a_m}}{\sum_{j\in\mathcal E}e^{a_j}},\qquad
m_*=\operatorname*{arg\,max}_{m\in\mathcal E}w_m.
\label{eq:attention}
\end{align}
where $a_m^{\rm phys}$ summarizes the numerical test results, $g_{\phi_s}$ is a learnable attention-based scoring function with parameters $\phi_s$. It adjusts each candidate score using the encoded operating context and the probe-derived features of all eligible candidates. If $\mathcal E$ is empty, ordinary numerical continuation is used.

This context helps distinguish similar test residuals
under different loading patterns. A historical segment
appropriate for load growth near one voltage-sensitive
area may be less useful when the additional demand
moves elsewhere. Recent accepted states also indicate
changes in the present trajectory direction.
The weights select one guide; numerical correction
determines the accepted PF states.

Offline training evaluates candidate-guided
continuation from each training state.
Let $\mathcal R\subseteq\mathcal E$ contain candidates
meeting the prescribed branch, accuracy, completion,
and subsequent-work criteria. For
$\mathcal R\ne\varnothing$, the loss is $\ell=-\log(\sum_{m\in\mathcal R}w_m).$
Several segments can support reliable continuation,
so training rewards the probability assigned to their
set. Labels depend on corrected trajectories and
subsequent work. States with no reliable candidate
are retained for coverage and fallback evaluation.
Future states used to construct offline labels
are unavailable to the online encoder. The encoder parameters are learned offline and held fixed during online evaluation.

\subsection{Guided Continuation and Loading-Limit Refinement}

For $u\in[0,1]$, the selected endpoint and ending tangent define a cubic Hermite guide satisfying
\begin{equation}
\begin{aligned}
\bm z_g(0)=\bm z_0, \
\bm z_g(1)=\widehat{\bm z}_{m_*}, \
\bm z_g'(0)=h\bm t_0, \
\bm z_g'(1)=h\widehat{\bm t}_{m_*}.
\end{aligned}
\label{eq:guide}
\end{equation}
The guide matches the starting and predicted ending states and their directions. It supplies intermediate predictions over the transferred segment. The interpolation parameter $u$ is not an exact arc-length coordinate on the present PF curve.

Starting from the last accepted state, the numerical predictor--corrector procedure uses an intermediate guide point as its prediction and corrects it with the present PF equations and a local continuation plane of the form in \eqref{eq:probe}. A state is accepted only when PF and continuation tolerances are satisfied. Its unit tangent is then recomputed from \eqref{eq:tangent} and oriented consistently with the preceding accepted tangent. Correction behavior and agreement with the proposed direction indicate whether the guide remains useful.

The requested advance $h$ specifies the extent of historical shape being reused. Numerical step sizes determine the intermediate corrections needed to follow the present curve. Useful guidance can reduce their number; unsuitable guidance causes step reduction or a return to the ordinary CPF predictor. Thus a successful endpoint test does not eliminate intermediate branch-tracking checks.

Progress is measured along the PF curve rather than by requiring $\lambda$ to increase at every step. Near a nose, a positive-to-negative change in $d\lambda/ds$ brackets a possible loading maximum. Within the associated neighborhood, final refinement solves
\begin{equation} \bm G(\bm x,\lambda,\bm v)= \begin{bmatrix} \bm F(\bm x,\lambda)\\ \bm J(\bm x,\lambda)\bm v\\ \bm b^{\mathsf T}\bm v-1 \end{bmatrix} =\bm0, \label{eq:fold} \end{equation}
where $\bm v$ is the critical right-null vector of the ordinary PF Jacobian and fixed $\bm b$ provides a nonzero normalization of that direction. The refined fold must be associated with the tracked branch. Identifying it as the first loading maximum additionally requires checking that no earlier maximum was skipped. The present formulation focuses on smooth PF segments with fixed bus types and control equations.

Algorithm~\ref{alg:trajectory_reuse} summarizes
the online loading-limit assessment.
Historical guidance is beneficial when reduced
continuation work exceeds retrieval, encoding,
candidate testing, and selection costs.
All online tests, rejected corrections, tangent
updates, and fallback operations enter the
computational comparison.

\begin{algorithm}[t]
\caption{Trajectory-Guided Loadability Assessment}
\label{alg:trajectory_reuse}
\begin{algorithmic}[1]
\REQUIRE Present PF model, historical segments,
requested advance $h$, computational budget
\ENSURE Refined loading limit $\lambda_{\rm lim}$
or failure
\STATE Initialize PF state, tangent, and accepted-state history
\WHILE{budget remains and the first fold is not bracketed}
    \STATE Retrieve seven historical candidates
    \STATE Align candidates using
    \eqref{eq:align}--\eqref{eq:endpoint}
    \STATE Test each candidate via Newton
    iterations in \eqref{eq:probe}
    \IF{eligible candidates exist}
        \STATE Select guidance using \eqref{eq:attention}
        and construct \eqref{eq:guide}
    \ELSE
        \STATE Use the ordinary CPF predictor
    \ENDIF
    \STATE Perform guarded PF corrections
    \STATE Update tangents and history for accepted states
    \STATE Check for the first positive-to-negative
    reversal of $d\lambda/ds$
\ENDWHILE
\IF{the first loading maximum is bracketed}
    \STATE Refine via \eqref{eq:fold} and verify branch
    \STATE Return $\lambda_{\rm lim}$ if accepted;
    otherwise report failure
\ELSE
    \STATE Report failure
\ENDIF
\end{algorithmic}
\end{algorithm}

\section{Theoretical Guarantees}
\label{sec:theory}

The analysis relates historical shape compatibility
to prediction error, local PF correction, and
computational benefit. All states use
\eqref{eq:scaled-state}. The results
apply to smooth PF segments between control events.

\subsection{Prediction Accuracy and Local Branch Preservation}

Let $\bm z(s)$ denote the present PF curve over $0\leq s\leq h$. The aligned historical prediction $\widetilde{\bm z}(s)$ matches its starting state and tangent. Define the difference in curve bending as
\begin{equation} \delta= \sup_{0\leq s\leq h} \left\| \widetilde{\bm z}''(s)-\bm z''(s) \right\|_2. \label{eq:compatibility} \end{equation}

\begin{proposition}[Historical-segment prediction error]
\label{thm:transfer}
Suppose the present and aligned historical curves are twice continuously differentiable, with $\widetilde{\bm z}(0)=\bm z(0)$ and $\widetilde{\bm z}'(0)=\bm z'(0)$. Then
\begin{align} \|\widetilde{\bm z}(h)-\bm z(h)\|_2 &\leq \delta h^2/2, \label{eq:prediction-bound}\\ \|\widetilde{\bm z}'(h)-\bm z'(h)\|_2 &\leq \delta h. \label{eq:direction-bound} \end{align}
\end{proposition}

\begin{proof}
The starting state and tangent differences vanish,
so
\[
 \widetilde{\bm z}(h)-\bm z(h)
 =
 \int_0^h(h-s)
 [\widetilde{\bm z}''(s)-\bm z''(s)]\,ds.
\]
Taking norms proves \eqref{eq:prediction-bound};
a single integration gives \eqref{eq:direction-bound}.
\end{proof}

The CPF tangent predictor has the corresponding
error bound $Mh^2/2$, where
$M=\sup_{0\leq s\leq h}\|\bm z''(s)\|_2$.
It neglects curve bending, whereas historical
transfer depends on the difference between historical
and present bending. Thus $\delta<M$ gives a
tighter bound over the same advance.
The potential benefit is a smaller error coefficient,
including on strongly curved P--V segments;
it is conditional on historical compatibility.

To connect this prediction to power balance,
suppose $\|\bm J_a\|_2\leq K_F$ along the line
joining $\widehat{\bm z}$ and $\bm z(h)$. Then
\begin{equation}
 \|\bm F(\widehat{\bm z})\|_2
 \leq K_F\|\widehat{\bm z}-\bm z(h)\|_2
 \leq K_F\delta h^2/2.
 \label{eq:residual-bound}
\end{equation}
Let $\bm H$ stack the PF equations and continuation
plane in \eqref{eq:probe}, and assume
$\bm B=\partial\bm H/\partial\bm z$ at
$\widehat{\bm z}$ is nonsingular.
Set $\beta=\|\bm B^{-1}\|_2$ and
$\eta=\|\bm B^{-1}\bm H(\widehat{\bm z})\|_2$.
If the Jacobian of $\bm H$ is Lipschitz continuous
with constant $L$ on a convex neighborhood containing
the closed radius-$2\eta$ ball centered at
$\widehat{\bm z}$, then $\beta L\eta<1/2$
gives local Newton convergence \cite{ortega1970},
with
\begin{equation}
 \|\bm z^\dagger-\widehat{\bm z}\|_2
 \leq2\eta\leq\beta K_F\delta h^2.
 \label{eq:correction-bound}
\end{equation}

\begin{proposition}[Local branch preservation]
\label{prop:branch}
Suppose $\bm J_a$ has full row rank at $\bm z(h)$.
Let the ball of radius $\rho_{\rm br}>0$ centered
at $\bm z(h)$ contain PF solutions only from the
intended local branch. If the preceding Newton
conditions hold and
\begin{equation}
 \delta h<1,\qquad
 \left(\beta K_F+\frac12\right)\delta h^2
 <\rho_{\rm br},
 \label{eq:branch-condition}
\end{equation}
then the correction converges to the intended local PF branch, with a continuation plane transverse to the tangent at $\bm z(h)$.
\end{proposition}

\begin{proof}
By \eqref{eq:direction-bound} and the unit-tangent property,
\begin{equation}
\widehat{\bm t}^{\mathsf T}\bm z'(h)
\geq 1-\delta h>0.
\label{eq:branch-direction}
\end{equation}
Thus, the continuation plane is transverse to the target branch. Moreover, combining \eqref{eq:prediction-bound} and \eqref{eq:correction-bound} gives
\begin{equation}
\begin{aligned}
\|\bm z^\dagger-\bm z(h)\|_2
&\leq\|\bm z^\dagger-\widehat{\bm z}\|_2
+\|\widehat{\bm z}-\bm z(h)\|_2\\
&\leq\left(\beta K_F+\frac12\right)\delta h^2
<\rho_{\rm br}.
\end{aligned}
\label{eq:branch-preservation}
\end{equation}
Since $\bm z^\dagger$ satisfies the PF equations and lies within the specified local neighborhood, it belongs to the intended branch.
\end{proof}

The first Newton test measures $\eta$, while
mismatch reduction and directional agreement
provide evidence of correctability.
Near a simple nose, $\bm J_a$ can retain full row
rank despite singularity of the ordinary PF Jacobian.
The results are local: $\delta$ and $\rho_{\rm br}$
are not known online, and the corrected section
solution need not lie at exactly arc length $h$.
Intermediate branch tracking remains necessary.
Once the associated simple fold is reached,
a nonsingular fold-system Jacobian gives a local
loading-limit error bound
$|\widehat\lambda_{\rm lim}-\lambda_{\rm lim}|
\leq C_{\rm fold}\tau_{\rm fold}$, where \(\tau_{\mathrm{fold}}\) bounds the final fold-system residual and \(C_{\mathrm{fold}}\) is a local conditioning constant.
Final accuracy follows from present-equation
refinement; identifying the first nose additionally
requires checking that no earlier maximum was skipped.

\subsection{Candidate Selection and Computational Benefit}

A historical segment is useful when it reduces
the work of reaching the specified continuation
solution, rather than merely resembling the initial
voltage profile. Two limitations are distinguished:
the retrieved group may contain no useful segment,
or the selector may overlook a useful segment
that is available. Alignment affects candidate
quality; physical testing and learned ranking
affect selection.

Let $C_m$ denote subsequent continuation work
using candidate $m$, including recovery and
fallback to reach the same target at the same
tolerances. For nonempty $\mathcal E$, define
$C_*=\min_{m\in\mathcal E}C_m$.
Let $C_{\rm test}$ include retrieval, encoding,
all Newton tests, and selection, measured using
a common computational metric.
Improvement over baseline CPF work $C_{\rm CPF}$
requires $
 C_{\rm CPF}-C_*
 >
 C_{\rm test}+(C_{m_*}-C_*).
 \label{eq:work-condition}$
The left side is the saving available from the
best candidate. The right side includes selection
overhead and extra work from imperfect selection.
This comparison explains why better prediction
alone does not establish lower total computation.

\section{Experimental Results}
\label{sec:experimental_results}

\subsection{Test Systems and Evaluation Protocol}

\begin{figure}[t]
    \centering
    \includegraphics[width=0.85\linewidth]
    {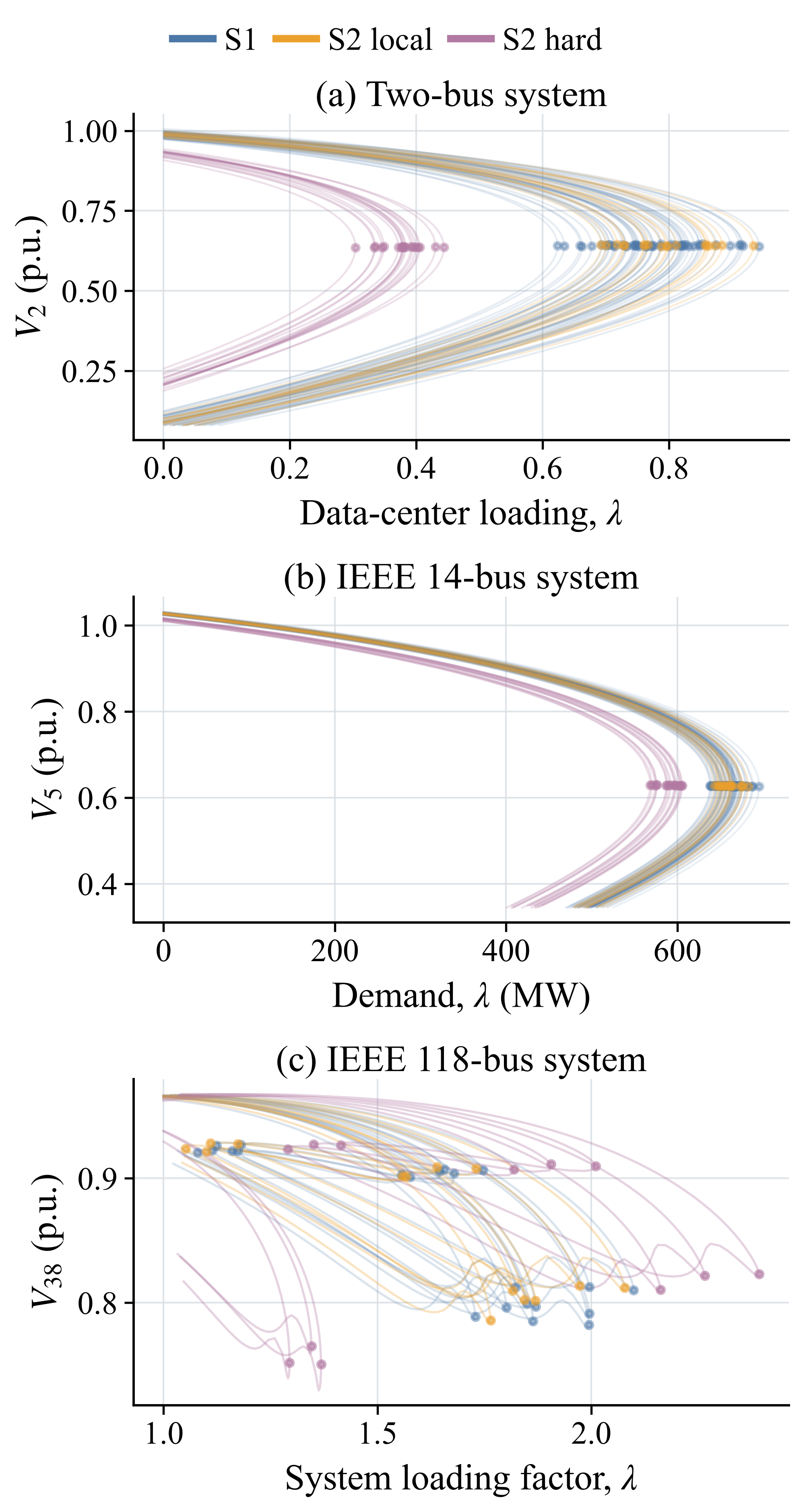}
    \caption{Historical S1 and held-out S2 P--V curves.
    Changes in generation and loading patterns alter
    voltage profiles, loading limits, and trajectory shape.}
    \label{fig:s1_s2_pv_families}
\end{figure}

We evaluate the proposed method on a two-bus system
adapted from the classical voltage-stability model
\cite{Kundur1994Power} and modified IEEE 14- and
118-bus systems \cite{Zimmerman2011MATPOWER}.
The two-bus system illustrates trajectory transfer
near a simple nose. The IEEE 14-bus system supports
the main continuation and component comparisons and the IEEE 118-bus system examines performance
on a larger network. In the IEEE 14-bus system, a fixed-power-factor
data center is connected at bus~5, solar generation
at buses~9 and~14, and wind generation at buses~12
and~13. The IEEE 118-bus configuration approximately
preserves this asset density, with eight data-center
buses, 17 solar buses, 17 wind buses, and eight
reactive-support buses. The aggregate solar and wind nameplate capacities are each set to 20\% of the base active demand and the total data center load demand varies by scenario.

S1 contains historical operating scenarios.
S2-local scenarios apply small perturbations
to corresponding S1 conditions. S2-hard includes
two separated regimes: low renewable availability with
higher reactive demand, larger data-center loads,
and limited reactive support; and high renewable
generation with light demand.
Figure~\ref{fig:s1_s2_pv_families} shows the resulting
P--V curves. In the representative IEEE 118-bus
trajectory, repeated sign changes in
$\mathrm{d}\lambda/\mathrm{d}s$ produce multiple
turns in the $(\lambda,V_{38})$ projection.
Near-singularity of the PF Jacobian and persistence
under step refinement support their interpretation
as folds of the PF solution curve.

Complete scenarios are separated into training,
validation, and test sets before normalization
and historical-database construction. Two tasks
are evaluated separately. The continuation-target
task asks whether a method reaches a specified
solution along the tracked branch, including
advances across a nose. The loading-limit task
terminates at the first loading maximum reached
from the common starting solution.
The primary IEEE 14-bus target-recovery population
contains 347 predeclared queries: 176 S2-local
and 171 S2-hard. Each specifies a corrected starting
state, forward direction, previously accepted states,
and requested arc-length advance. Future evaluation
states are unavailable to the predictor.

For target recovery, all methods are evaluated
against the same prescribed reference
$\bm z^\star$. Success requires at least 96\%
of the requested advance, final PF residual
not exceeding $10^{-8}$, normalized refined-target
error not exceeding $0.10$, and positive terminal
orientation toward $\bm z^\star$.
PF feasibility alone is therefore insufficient.
Feasible completions violating the target-error
or orientation criterion are classified as
wrong-target convergence. Numerical breakdown
and termination before the required advance
are reported separately. Time and work budgets
are checked after complete solver calls.

Adaptive CPF is the baseline for continuation-target
recovery. Loading-limit assessment additionally
includes HEAP \cite{Zhang2022HEAP} and the fixed-active-set
curve-construction core E-HEAP
\cite{Zhang2025RobustHE}. Within each experiment,
the methods share the initialization, present PF
equations, external acceptance criteria, and
computational environment. Reported online costs
include candidate testing, correction, tangent
updates, rejected attempts, and fallback.
Offline training and historical-database construction
are excluded, so the timing results characterize
online assessment rather than amortized total cost.

The proposed method uses four accepted continuation
states and seven aligned historical candidates.
Each candidate receives up to three Newton test
iterations, with early termination when screening
criteria fail. Adaptive CPF uses a validation-tuned
step controller and a nominal step of $0.05$
in the loading-limit study. HEAP and E-HEAP core
use maximum orders of 12 and 6, with internal
mismatch tolerances of $10^{-5}$ and $10^{-6}$,
respectively. These internal settings are distinct
from the common final acceptance checks.

Experiments use an Intel Core Ultra 7 251HX processor
and an NVIDIA GeForce RTX 5060 Laptop GPU with
8~GB memory, running Python, PyTorch 2.12.1+cu130,
and CUDA 13.0. Repeated target-recovery timing uses
32 fixed queries per split, three repetitions per
method and query, and eight warm-up queries.
Methods alternate execution order, with batch size
one, memory-resident models and historical segments,
and GPU synchronization before and after each
timed call.

\subsection{Continuation-Target Recovery Near P--V Noses}

Figures~\ref{fig:two_bus_comparison}--
\ref{fig:ieee118_comparison} compare recovery
of specified continuation solutions.
The three-dimensional views display the tracked
trajectories, while the two-dimensional projections
locate unsuccessful endpoints relative to the noses.
Crosses denote endpoints failing the common recovery
criterion. These displayed cases are reported
separately from the aggregate IEEE 14-bus population
in Table~\ref{tab:end_to_end_online_performance}.

\begin{figure}[t]
    \centering
    \includegraphics[width=0.95\linewidth]
    {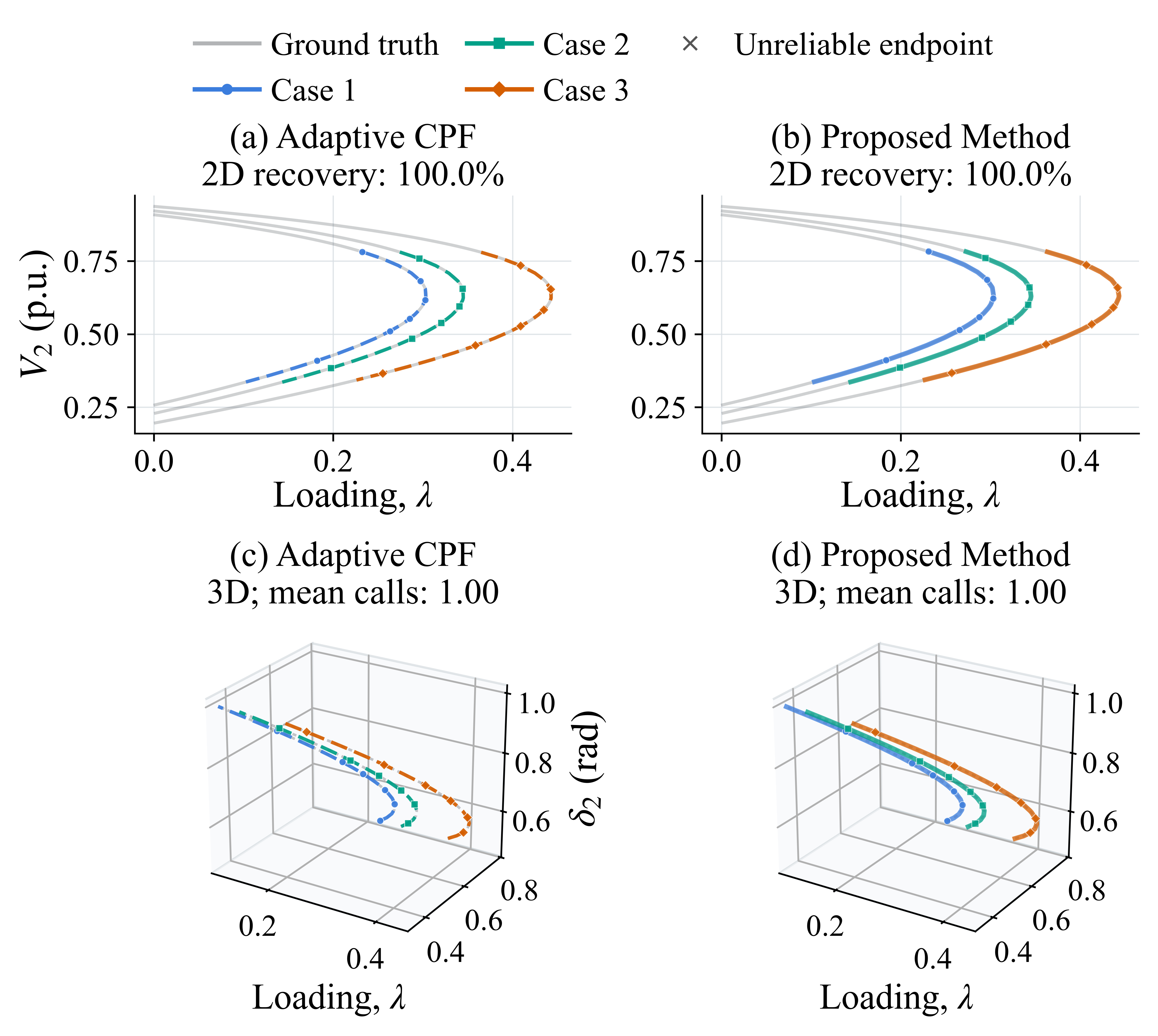}
    \caption{Two-bus continuation results:
    Adaptive CPF and the proposed method.}
    \label{fig:two_bus_comparison}
\end{figure}

\begin{figure}[t]
    \centering
    \includegraphics[width=0.95\linewidth]
    {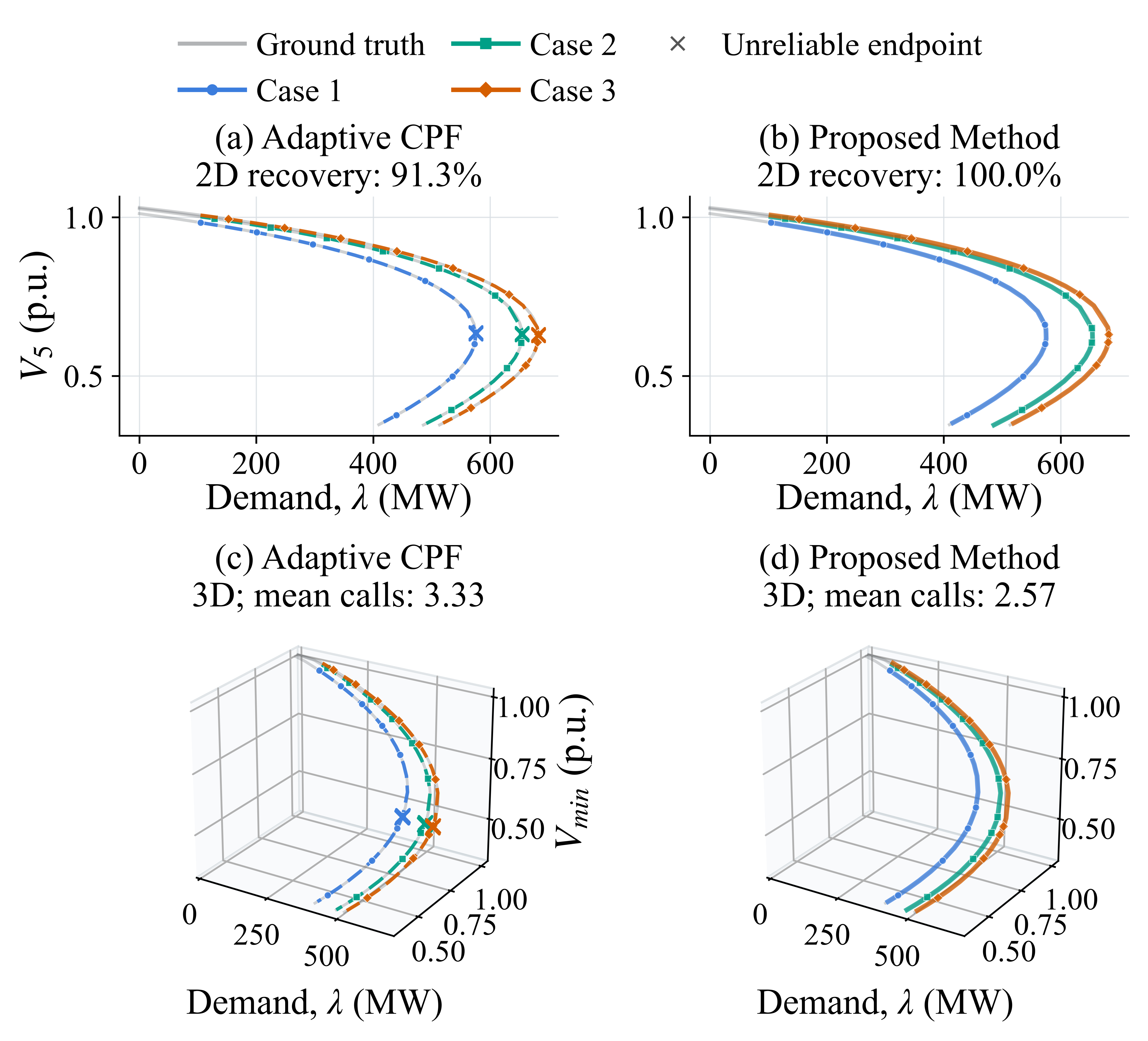}
    \caption{Continuation-target recovery on the modified IEEE 14-bus system. Crosses denote unsuccessful endpoints.}
    \label{fig:ieee14_comparison}
\end{figure}

\begin{figure}[t]
    \centering
    \includegraphics[width=0.95\linewidth]
    {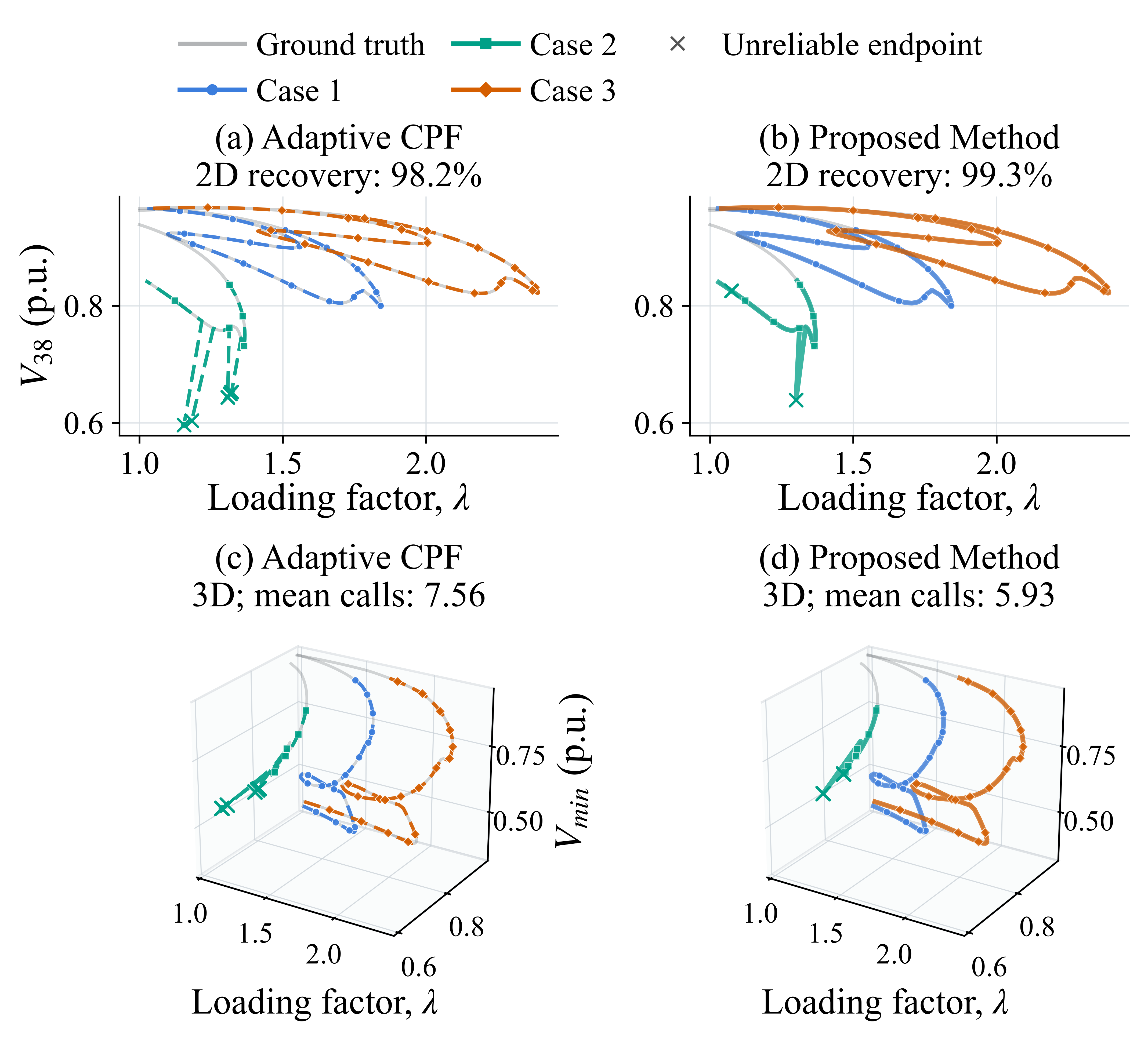}
    \caption{Continuation-target recovery on the modified IEEE 118-bus system. Crosses denote unsuccessful endpoints.}
    \label{fig:ieee118_comparison}
\end{figure}

Both methods recover all displayed two-bus targets
with one call to the continuation routine, showing
that the simple single-fold case is readily handled
by conventional continuation.
For the displayed IEEE 14-bus cases, Adaptive CPF
recovers 105 of 115 targets, whereas the proposed
method recovers all 115. Mean continuation-routine
calls decrease from 3.33 to 2.57.
For the IEEE 118-bus cases, Adaptive CPF recovers
274 of 279 targets with 7.56 mean calls.
The proposed method recovers 277 targets with
5.93 mean calls, a 21.6\% reduction.
Routine-call counts describe execution behavior;
total linear solves provide the more detailed
work comparison below.

\begin{figure}[t]
    \centering
    \includegraphics[width=\linewidth]
    {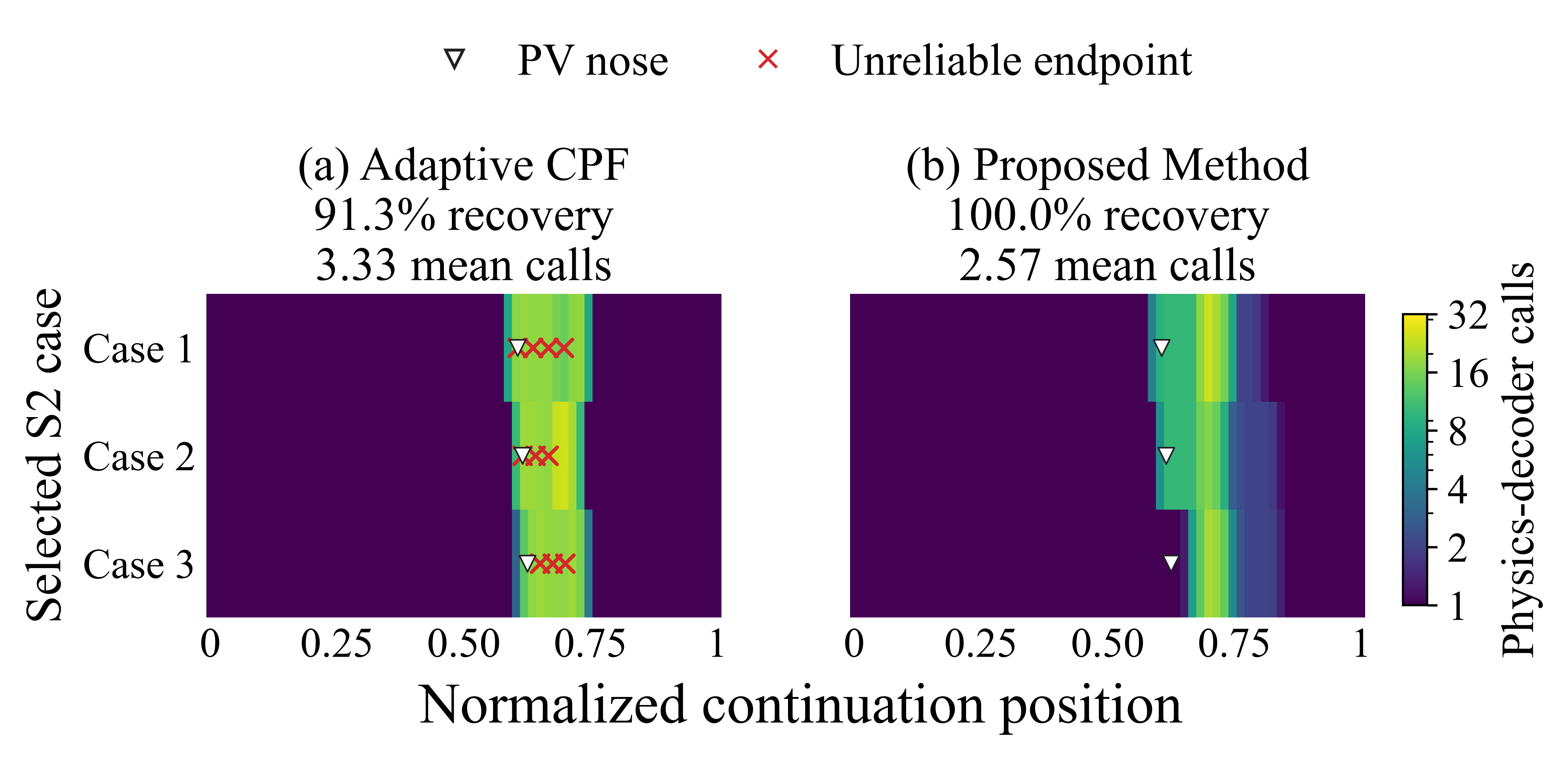}
    \caption{Continuation-routine calls and unsuccessful
    endpoints along three displayed IEEE 14-bus
    continuations.}
    \label{fig:ieee14_branch_risk}
\end{figure}

All ten unsuccessful Adaptive CPF queries in the
displayed IEEE 14-bus cases occur at or immediately
beyond the nose and stop before completing the required
advance. An audit identifies 103 rejected trials: 102 fail
Newton convergence or the residual criterion, and
one violates the correction-ratio bound.
Repeated rejection reduces the trial step below
its prescribed minimum.

For the difficult Case~1 query at target index~110,
Adaptive CPF accepts four trials and rejects ten
before minimum-step termination. The proposed method
selects an eligible aligned historical candidate
and completes 96.6\% of the requested advance,
with normalized target error 0.034, PF residual
$4.18\times10^{-14}$, and positive terminal orientation.
This example illustrates how historical shape can
help continue through a region where repeated local
predictions lead to correction failure and step
contraction. It also shows why PF residual and
completed advancement must be evaluated together.

\subsection{Effects of Historical Reuse and Candidate Selection}

\begin{figure}[t]
    \centering
    \includegraphics[width=0.8\columnwidth]
    {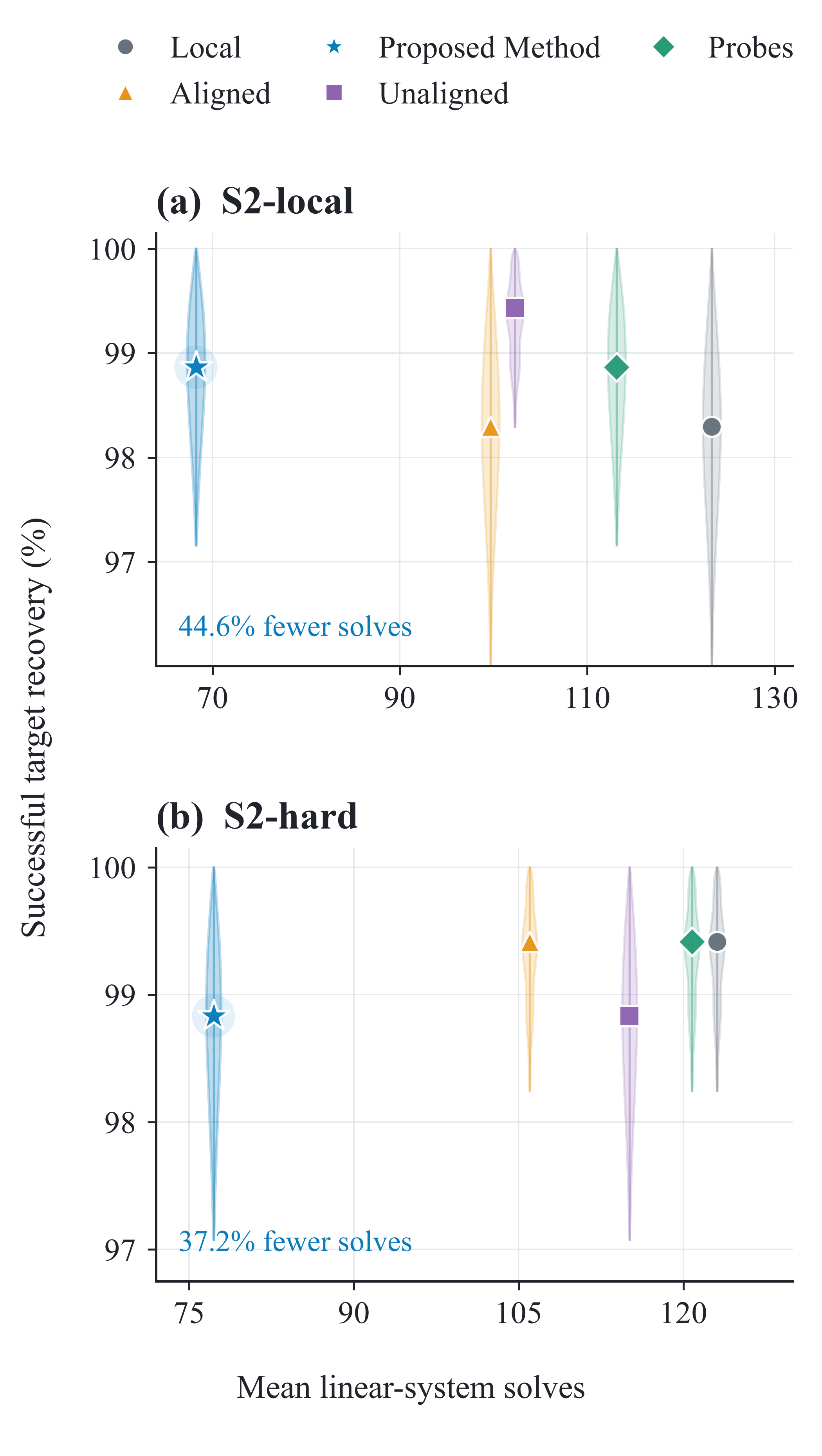}
    \caption{Effects of historical reuse, alignment,
    and candidate selection. Markers show recovery
    rates and mean linear-system solves; violins show
    the central 95\% bootstrap distributions.}
    \label{fig:source_trajectory_reuse_effectiveness}
\end{figure}

Figure~\ref{fig:source_trajectory_reuse_effectiveness}
examines local prediction, unaligned historical reuse,
aligned reuse, reuse with Newton tests, and the full
proposed method. The configurations use the same
176 S2-local and 171 S2-hard queries, external
acceptance criterion, and guarded continuation
controller. Work includes candidate tests, tangent
updates, rejected corrections, and fallback.

Relative to local prediction, unaligned reuse reduces
mean work by 17.0\% on S2-local and 6.5\% on S2-hard,
with paired 95\% confidence intervals of
$[13.0,21.4]\%$ and $[4.4,8.6]\%$.
The recovery-rate differences are small, with
confidence intervals including zero.
Alignment further reduces work by 2.5\% and 7.9\%,
respectively. Historical segments therefore provide
useful computational information, and matching their
position and direction improves its use.

Newton testing alone does not establish a
population-level recovery improvement. Compared
with aligned reuse without tests, always-on testing
increases S2-local recovery by 0.57 percentage points,
with a 95\% confidence interval of
$[-1.14,2.84]$ percentage points, and produces no
change on S2-hard. Mean work increases by 13.5\%
and 13.9\%, respectively. Candidate validation thus
has a measurable cost that must be recovered through
better subsequent continuation.

The full method retains testing of all seven
candidates and combines the resulting evidence
with operating-condition information for selection.
Relative to reuse with Newton tests, it reduces
mean work by 39.7\% on S2-local and 36.1\% on
S2-hard, with recovery rates close to 99\%.
The observed benefit concerns the complete guidance
selection and continuation procedure; these results
do not separately establish a reliability gain
from attention or Newton testing alone.
They support the design principle that historical
information should be evaluated by its usefulness
for subsequent PF computation.

\subsection{Recovery Reliability and Computational Effort}

\begin{table}[t]
\centering
\caption{IEEE 14-bus continuation-target recovery
and online computation. Work entries are means over
all queries; runtime uses 32 queries per split.
P95 denotes the 95th percentile and CI denotes
confidence interval.}
\label{tab:end_to_end_online_performance}
\resizebox{\columnwidth}{!}{%
\setlength{\tabcolsep}{3.5pt}
\begin{tabular}{llcccc}
\toprule
Split & Method &
\makecell{Recovery\\(success/total)} &
\makecell{Corrector\\calls} &
\makecell{Test\\solves} &
\makecell{Total linear\\solves} \\
\midrule
S2-local & Adaptive CPF
& 97.73\% (172/176) & 15.63 & 0.00 & 158.73 \\
& Proposed
& 99.43\% (175/176) & 7.14 & 11.81 & 67.49 \\
S2-hard & Adaptive CPF
& 98.25\% (168/171) & 16.32 & 0.00 & 171.09 \\
& Proposed
& 99.42\% (170/171) & 7.74 & 12.65 & 75.87 \\
\midrule
Split & Method &
\makecell{Median/P95\\runtime (ms)} &
\multicolumn{2}{c}{
\makecell{Paired geometric speedup\\
(95\% bootstrap CI)}} &
\makecell{Proposed faster\\fraction} \\
\midrule
S2-local & Adaptive CPF
& 770.38/909.65
& \multicolumn{2}{c}{$1.00\times$} & -- \\
& Proposed
& 281.95/910.27
& \multicolumn{2}{c}{$5.42\times$ [2.89, 10.20]}
& 87.5\% \\
S2-hard & Adaptive CPF
& 671.02/821.68
& \multicolumn{2}{c}{$1.00\times$} & -- \\
& Proposed
& 80.28/560.83
& \multicolumn{2}{c}{$5.85\times$ [3.25, 10.41]}
& 90.6\% \\
\bottomrule
\end{tabular}}
\end{table}

Table~\ref{tab:end_to_end_online_performance}
reports all 347 primary IEEE 14-bus queries,
including unsuccessful terminations.
Corrector calls include accepted, rejected, and
fallback attempts. Total work includes correction,
candidate-testing, and tangent linear systems.
Observed recovery increases from 97.73\% to 99.43\%
on S2-local and from 98.25\% to 99.42\% on S2-hard.
Mean total solves decrease from 158.73 to 67.49
and from 171.09 to 75.87, respectively.
The work reduction is the principal computational
result; the recovery percentages describe observed
outcomes under the evaluated settings.

Adaptive CPF records four S2-local and three S2-hard
budget-exhaustion outcomes. The proposed method
records one wrong-target termination on S2-local
and one budget exhaustion on S2-hard.
Neither reports explicit numerical breakdown.
Because rejected corrections and repeated step
reductions consume the available work, budget
exhaustion can reflect numerical difficulty near
a nose as well as the imposed computational limit.

\begin{figure}[t]
    \centering
    \includegraphics[width=0.8\columnwidth]
    {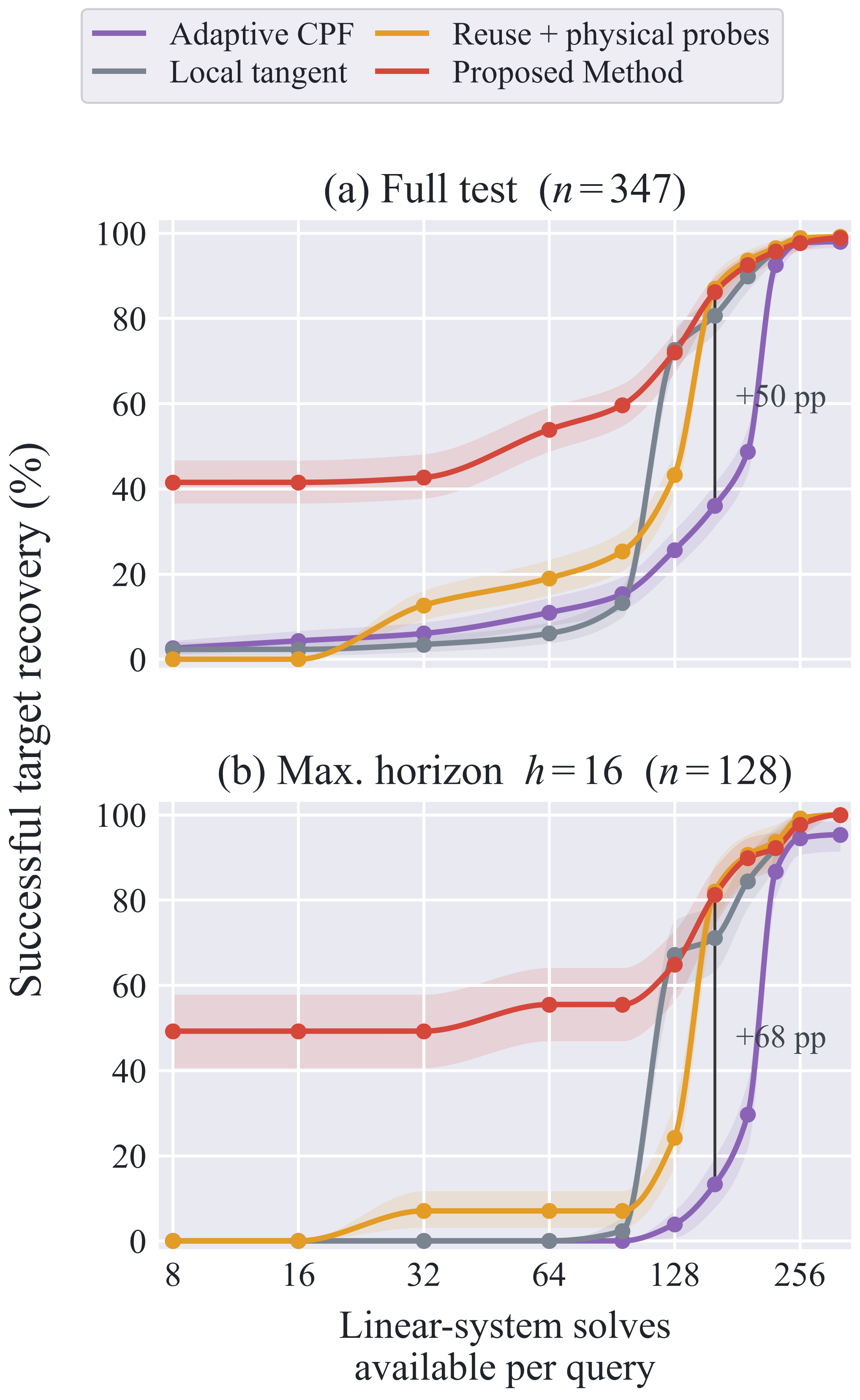}
    \caption{Recovery of specified continuation
    solutions versus nominal linear-system-solve
    budget, for all 347 queries and the
    maximum-advance subset.}
    \label{fig:reliability_primary_cohorts}
\end{figure}

Figure~\ref{fig:reliability_primary_cohorts} shows
the recovery--work relationship.
Near a nominal budget of 160 linear solves, the
proposed method has approximately 50 percentage
points higher recovery over all queries and
68 percentage points higher recovery over the
128 maximum-advance queries. Adaptive CPF approaches
the proposed method as the budget increases.
The comparison indicates that historical guidance
can reduce the work needed for long continuation
advances, rather than establishing that the
baseline cannot recover those solutions.
Since budgets are checked after complete calls,
the horizontal axis represents nominal allowances.

Runtime statistics use the median of three
repetitions for each method and query.
Paired geometric-mean speedups are computed from
query-level runtime ratios, with 95\% confidence
intervals from 10,000 paired bootstrap resamples.
All outcomes are retained, including time spent
before unsuccessful termination.

Median runtime decreases from 770.38 to 281.95~ms
on S2-local and from 671.02 to 80.28~ms on S2-hard.
The paired geometric-mean speedups are
$5.42\times$ [2.89, 10.20] and
$5.85\times$ [3.25, 10.41], respectively.
These paired statistics differ from ratios of
cohort medians. The proposed method is faster
on 87.5\% and 90.6\% of timed queries.
However, its S2-local P95 is comparable to
Adaptive CPF, reflecting costly guarded or fallback
execution in difficult cases. The evidence therefore
supports lower typical online computation,
with remaining variability across operating cases.

\subsection{Voltage-Stability Loading-Limit Assessment}
\label{subsec:updated_loadability}

The preceding experiments evaluate prescribed
continuation targets, including states beyond a nose.
For large-load assessment, the required endpoint
is instead the first local loading maximum reached
from the corrected high-voltage solution.
This experiment evaluates that operational task
directly. The proposed method uses historical
segments to approach the critical region, while
present-case PF correction determines accepted states.
Assessment terminates after a physical loading-tangent
sign change brackets the first nose and the common
fold-refinement procedure converges.

For each S2 scenario,
$\lambda_{\mathrm{lim}}^{\mathrm{ref}}$ is the
independently refined reference limit and
$\widehat{\lambda}_{\mathrm{lim}}$ is the assessed
limit. The acceptance checks use $e_{\mathrm{lim}}
=
\left|
\widehat{\lambda}_{\mathrm{lim}}
-\lambda_{\mathrm{lim}}^{\mathrm{ref}}
\right|, \
e_{\mathrm{over}}=
\max\left\{
0,\widehat{\lambda}_{\mathrm{lim}}
-\lambda_{\mathrm{lim}}^{\mathrm{ref}}
\right\},$
converted to MW using the prescribed data-center
demand increase per unit $\lambda$.
The stored S2 curve is used offline to identify
the reference fold neighborhood. Online methods
receive only the common corrected starting prefix;
future S2 states and the reference limit are
unavailable. Acceptance requires the intended branch,
a bracketed and refined fold, and the common PF
residual and loading-limit error checks.


\begin{table*}[t]
\centering
\caption{Online loading-limit assessment on modified
IEEE 14- and 118-bus systems. Times are median (P95)
over the scenarios in each row.}
\label{tab:updated_loadability_all_methods}
\footnotesize
\setlength{\tabcolsep}{3.8pt}
\renewcommand{\arraystretch}{1.08}
\begin{tabular}{@{}llc@{\hspace{5pt}}cccc@{\hspace{5pt}}ccc@{}}
\toprule
& & Valid &
\multicolumn{4}{c}{Online time (s)} &
\multicolumn{3}{c}{Proposed paired speedup} \\
\cmidrule(lr){4-7}
\cmidrule(lr){8-10}
System & Split & C/H/E/P &
CPF & HEAP & E-HEAP core & Proposed &
vs. CPF & vs. HEAP & vs. E-HEAP \\
\midrule
IEEE 14 & S2-local & 16/16/16/16
& 1.465 (2.032) & 0.423 (0.554)
& 0.410 (0.516) & \textbf{0.283 (0.338)}
& $4.88\times$ & $1.53\times$ & $1.41\times$ \\
& S2-hard & 16/16/15/16
& 1.383 (1.922) & 0.487 (0.539)
& 0.413 (0.473) & \textbf{0.255 (0.288)}
& $5.41\times$ & $1.86\times$ & $1.65\times$ \\
IEEE 118 & S2-local & 6/6/6/6
& 3.314 (3.533) & 0.763 (0.877)
& 3.177 (5.132) & \textbf{0.152 (0.277)}
& $21.62\times$ & $4.77\times$ & $22.03\times$ \\
& S2-hard & 6/6/6/6
& 2.755 (4.320) & 0.708 (0.783)
& 2.362 (3.278) & \textbf{0.185 (0.369)}
& $14.83\times$ & $3.48\times$ & $11.15\times$ \\
\bottomrule
\multicolumn{10}{l}{\scriptsize
C/H/E/P: Adaptive CPF/HEAP/E-HEAP core/Proposed.
Valid entries are accepted scenario counts.}\\
\multicolumn{10}{l}{\scriptsize
Speedups summarize paired scenario-wise runtime ratios,
rather than ratios of the displayed group medians.}
\end{tabular}
\end{table*}


\begin{figure}[t]
    \centering
    \includegraphics[width=0.85\columnwidth]
    {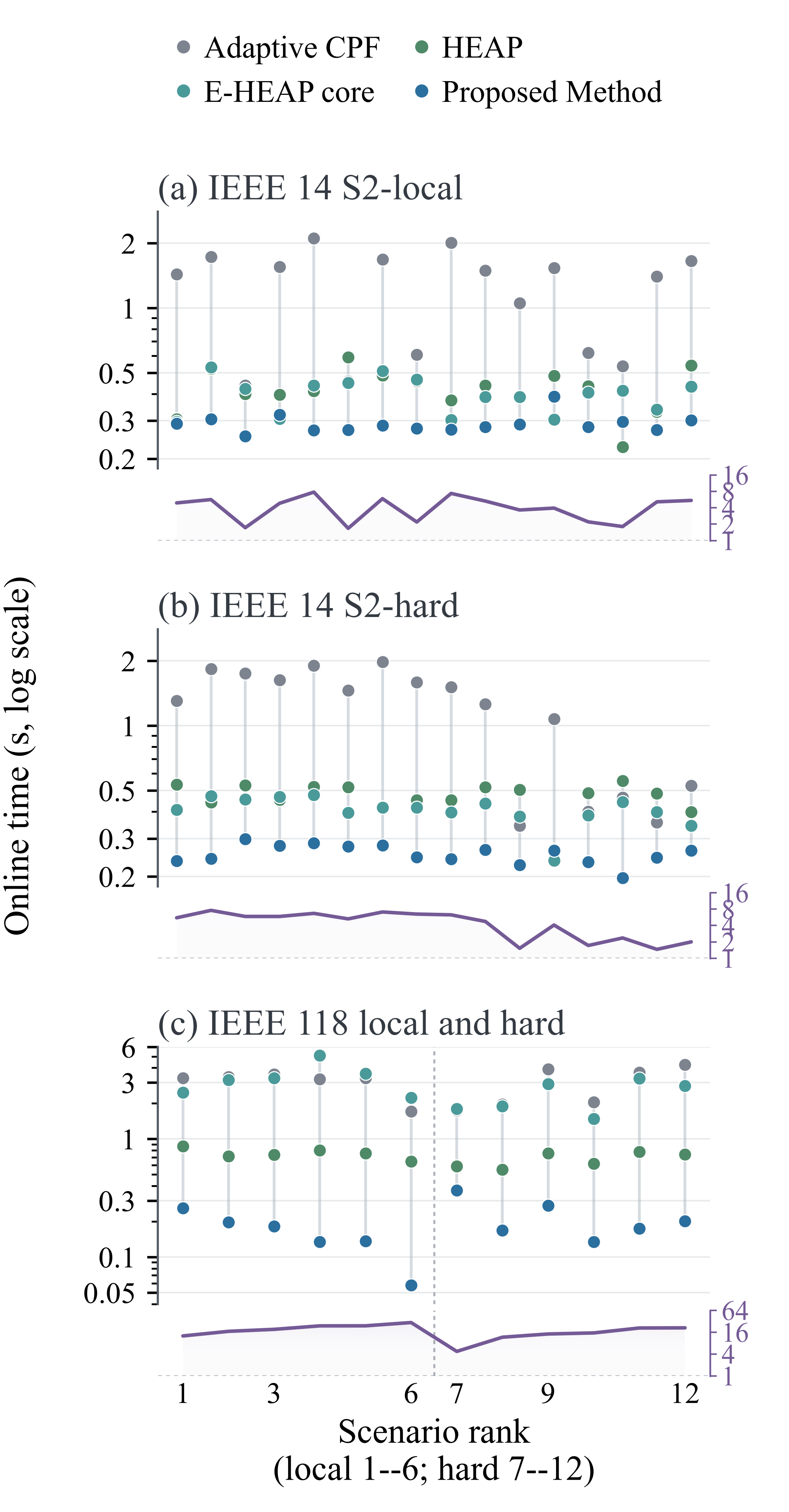}
    \caption{Scenario-wise online loading-limit
    assessment times and paired speedup of the
    proposed method over Adaptive CPF.}
    \label{fig:updated_loadability_all_methods}
\end{figure}

Adaptive CPF, HEAP \cite{Zhang2022HEAP}, and E-HEAP
core \cite{Zhang2025RobustHE} use the same corrected
starting prefix, terminal fold refinement, and
external acceptance criteria.
The E-HEAP comparison concerns its fixed-active-set
curve-construction core, consistent with the
fixed-control scope of this study.
A multidimensional holomorphic prototype based
on \cite{liu2017} is excluded from the timing ranking
because its truncated representation fails the
branch-acceptance checks under the evaluated large data-center shifts
near the fold. This prototype result is not
a general limitation of multidimensional
holomorphic embedding.

Table~\ref{tab:updated_loadability_all_methods}
reports 16 scenarios per split for IEEE 14
and six per split for IEEE 118.
The proposed method, Adaptive CPF, and HEAP pass
the acceptance checks in all 44 scenarios.
E-HEAP core passes 43, with one unsuccessful
IEEE 14-bus S2-hard assessment. For IEEE 14, the proposed method achieves paired
speedups of $4.88$--$5.41\times$ over Adaptive CPF,
$1.53$--$1.86\times$ over HEAP, and
$1.41$--$1.65\times$ over E-HEAP core.
For IEEE 118, the corresponding ranges are
$14.83$--$21.62\times$,
$3.48$--$4.77\times$, and
$11.15$--$22.03\times$.
The proposed method is faster than HEAP in all
12 IEEE 118-bus scenarios.
Across these 12 scenarios, the reported
paired-median speedups are $18.10\times$,
$3.85\times$, and $14.99\times$, respectively.
These online results are consistent with the
different sources of prediction information.
Adaptive CPF constructs the present curve through
successive local predictions and corrections.
HEAP and E-HEAP core construct present-case
holomorphic approximations. The proposed method
supplies previously computed finite-distance
trajectory shape and uses present-case solves
to test, correct, and refine that guidance.
Across the evaluated fixed-control scenarios,
this reuse reduces online assessment time while
meeting the common branch and loading-limit
acceptance checks. The results do not establish
an amortized advantage after offline preparation
costs or performance across control-limit events.

\section{Conclusion}
\label{sec:conclusion}

Repeated loadability studies for large-load integration must account for changing generation and demand patterns. Although related cases can exhibit similar P--V behavior, a historical operating point alone does not describe the subsequent voltage response, and transferring a trajectory without checking the present equations can misguide continuation. This paper addressed these challenges by aligning finite historical segments with the current state and tangent, thereby supplementing the CPF tangent predictor with previously computed nonlinear curve shape. Short Newton tests and attention-based selection identify electrically compatible guidance, while numerical correction and fold refinement determine the accepted PF solutions and loading limit. The analysis shows that prediction error depends on differences between historical and present curve bending, explaining when reuse can reduce the work of reconstructing a new trajectory. Its computational benefit requires those savings to exceed testing, selection, and recovery costs. The proposed approach provides a basis for reusing prior PF computations in repeated loading-margin assessments. Future work will evaluate larger networks, realistic large-load study sequences, and control-limit events.

\bibliographystyle{IEEEtran}
\bibliography{references}

@article{lundberg1991,
  author  = {B. N. Lundberg and A. B. Poore},
  title   = {Variable order {Adams--Bashforth} predictors with an error-stepsize control for continuation methods},
  journal = {SIAM Journal on Scientific and Statistical Computing},
  volume  = {12},
  number  = {3},
  pages   = {695--723},
  year    = {1991}
}

@article{liu2017,
  author  = {C. Liu and B. Wang and X. Xu and K. Sun and D. Shi and C. L. Bak},
  title   = {A multi-dimensional holomorphic embedding method to solve {AC} power flows},
  journal = {IEEE Access},
  volume  = {5},
  pages   = {25270--25285},
  year    = {2017}
}

@article{greene1997,
  author  = {S. Greene and I. Dobson and F. L. Alvarado},
  title   = {Sensitivity of the loading margin to voltage collapse with respect to arbitrary parameters},
  journal = {IEEE Transactions on Power Systems},
  volume  = {12},
  number  = {1},
  pages   = {262--272},
  year    = {1997}
}

@book{ortega1970,
  author    = {J. M. Ortega and W. C. Rheinboldt},
  title     = {Iterative Solution of Nonlinear Equations in Several Variables},
  publisher = {Academic Press},
  year      = {1970}
}

@article{Sun2023StaticVoltage,
  author  = {Y. Sun and T. Ding and O. Han and C. Liu and F. Li},
  title   = {Static voltage stability analysis based on multi-dimensional holomorphic embedding method},
  journal = {IEEE Transactions on Power Systems},
  volume  = {38},
  number  = {4},
  pages   = {3748--3759},
  month   = jul,
  year    = {2023}
}

@article{Wu2022TriSectional,
  author  = {D. Wu and B. Wang and F.-E. Wolter and L. Xie},
  title   = {Tri-sectional approximation of the shortest path to long-term voltage stability boundary with distributed energy resources},
  journal = {IEEE Transactions on Power Systems},
  volume  = {37},
  number  = {6},
  pages   = {4720--4731},
  month   = nov,
  year    = {2022}
}

@article{Luderer2022Electrification,
  author  = {G. Luderer and others},
  title   = {Impact of declining renewable energy costs on electrification in low-emission scenarios},
  journal = {Nature Energy},
  volume  = {7},
  number  = {1},
  pages   = {32--42},
  month   = jan,
  year    = {2022}
}

@article{Xiao2025AIServers,
  author  = {T. Xiao and others},
  title   = {Environmental impact and net-zero pathways for sustainable artificial intelligence servers in the {USA}},
  journal = {Nature Sustainability},
  volume  = {8},
  number  = {12},
  pages   = {1541--1553},
  month   = dec,
  year    = {2025}
}

@article{Gyang2026DataCenter,
  author  = {P. P. Gyang and P. Chakraborty and L. Meegahapola and X. Yu},
  title   = {Dynamic modeling of a data center for power system stability studies},
  journal = {IEEE Transactions on Power Systems},
  volume  = {41},
  number  = {3},
  pages   = {2317--2332},
  month   = may,
  year    = {2026}
}

@misc{PJM2026LargeLoad,
  author       = {{PJM Interconnection}},
  title        = {{PJM} proposes reliability standards to manage large load disconnection events},
  howpublished = {PJM Inside Lines},
  month        = sep,
  year         = {2026},
  url          = {https://insidelines.pjm.com/reliability-standards-to-manage-large-load-disconnection-events-proposed-by-pjm/},
  note         = {[Online; accessed 19-Sep-2026]}
}

@article{Ajjarapu1992CPF,
  author  = {V. Ajjarapu and C. Christy},
  title   = {The continuation power flow: A tool for steady state voltage stability analysis},
  journal = {IEEE Transactions on Power Systems},
  volume  = {7},
  number  = {1},
  pages   = {416--423},
  month   = feb,
  year    = {1992}
}

@inproceedings{DeepOPFAL2023,
  author    = {X. Pan and W. Huang and M. Chen and S. H. Low},
  title     = {{DeepOPF-AL}: Augmented learning for solving {AC-OPF} problems with a multi-valued load-solution mapping},
  booktitle = {Proceedings of the ACM International Conference on Future Energy Systems (e-Energy)},
  address   = {Orlando, FL, USA},
  pages     = {42--47},
  year      = {2023}
}

@article{DomainGeneralizationSurvey,
  author  = {J. Wang and C. Lan and C. Liu and Y. Ouyang and T. Qin and W. Lu and Y. Chen and W. Zeng and P. Yu},
  title   = {Generalizing to unseen domains: A survey on domain generalization},
  journal = {IEEE Transactions on Knowledge and Data Engineering},
  volume  = {35},
  number  = {8},
  pages   = {8052--8072},
  month   = aug,
  year    = {2023}
}

@article{Tinney1967Newton,
  author  = {W. F. Tinney and C. E. Hart},
  title   = {Power flow solution by {Newton}'s method},
  journal = {IEEE Transactions on Power Apparatus and Systems},
  volume  = {PAS-86},
  number  = {11},
  pages   = {1449--1460},
  month   = nov,
  year    = {1967}
}

@article{Stott1974FDLF,
  author  = {B. Stott and O. Alsac},
  title   = {Fast decoupled load flow},
  journal = {IEEE Transactions on Power Apparatus and Systems},
  volume  = {PAS-93},
  number  = {3},
  pages   = {859--869},
  month   = may,
  year    = {1974}
}

@article{Chiang1995CPFLOW,
  author  = {H.-D. Chiang and A. J. Flueck and K. S. Shah and N. Balu},
  title   = {{CPFLOW}: A practical tool for tracing power system steady-state stationary behavior due to load and generation variations},
  journal = {IEEE Transactions on Power Systems},
  volume  = {10},
  number  = {2},
  pages   = {623--634},
  month   = may,
  year    = {1995}
}

@article{Zhang2022HEAP,
  author  = {W. Zhang and T. Wang and H.-D. Chiang},
  title   = {A novel {FFHE}-inspired method for large power system static stability computation},
  journal = {IEEE Transactions on Power Systems},
  volume  = {37},
  number  = {1},
  pages   = {726--737},
  month   = jan,
  year    = {2022}
}

@article{Zhang2025RobustHE,
  author  = {W. Zhang and Y. Zhang and C. Zhang and H.-D. Chiang and T. Wang},
  title   = {A robust holomorphic embedding-based method for fast tracing of {P--V} curves with limiting points},
  journal = {IEEE Transactions on Power Systems},
  volume  = {40},
  number  = {6},
  pages   = {4688--4701},
  month   = nov,
  year    = {2025}
}

@article{Zeng2024ADMIRE,
  author  = {L. Zeng and H.-D. Chiang},
  title   = {Online and look-ahead determination of the renewable admissible region for managing the uncertainty of renewables: Theory and some applications},
  journal = {IEEE Transactions on Power Systems},
  volume  = {39},
  number  = {4},
  pages   = {5609--5619},
  month   = jul,
  year    = {2024}
}

@article{Zeng2026CoreX,
  author  = {L. Zeng and H.-D. Chiang},
  title   = {Online constructive method for computing renewable energy admissible injections region using adaptive expansion schemes: A new {EMS} function},
  journal = {IEEE Transactions on Power Systems},
  volume  = {41},
  number  = {2},
  pages   = {1011--1021},
  month   = mar,
  year    = {2026}
}

@article{Hu2021PhysicsGuidedPF,
  author  = {X. Hu and H. Hu and S. Verma and Z.-L. Zhang},
  title   = {Physics-guided deep neural networks for power flow analysis},
  journal = {IEEE Transactions on Power Systems},
  volume  = {36},
  number  = {3},
  pages   = {2082--2092},
  month   = may,
  year    = {2021}
}

@article{Yuan2022SupportMatrixPF,
  author  = {J. Yuan and Y. Weng},
  title   = {Support matrix regression for learning power flow in distribution grid with unobservability},
  journal = {IEEE Transactions on Power Systems},
  volume  = {37},
  number  = {2},
  pages   = {1151--1161},
  month   = mar,
  year    = {2022}
}

@article{Zhou2023DeepOPFFT,
  author  = {M. Zhou and M. Chen and S. H. Low},
  title   = {{DeepOPF-FT}: One deep neural network for multiple {AC-OPF} problems with flexible topology},
  journal = {IEEE Transactions on Power Systems},
  volume  = {38},
  number  = {1},
  pages   = {964--967},
  month   = jan,
  year    = {2023}
}

@article{Huang2024UnsupervisedOPF,
  author  = {W. Huang and M. Chen and S. H. Low},
  title   = {Unsupervised learning for solving {AC} optimal power flows: Design, analysis, and experiment},
  journal = {IEEE Transactions on Power Systems},
  volume  = {39},
  number  = {6},
  pages   = {7102--7114},
  month   = nov,
  year    = {2024}
}

@article{Zimmerman2011MATPOWER,
  author  = {R. D. Zimmerman and C. E. Murillo-S{\'a}nchez and R. J. Thomas},
  title   = {{MATPOWER}: Steady-state operations, planning, and analysis tools for power systems research and education},
  journal = {IEEE Transactions on Power Systems},
  volume  = {26},
  number  = {1},
  pages   = {12--19},
  month   = feb,
  year    = {2011}
}

@book{Kundur1994Power,
  author    = {P. Kundur},
  title     = {Power System Stability and Control},
  address   = {New York, NY, USA},
  publisher = {McGraw-Hill},
  year      = {1994}
}

\vfill

\end{document}